\documentclass{ws-ijmpcs}

\def\beq{\begin{equation}}
\def\eeq{\end{equation}}
\def\bea{\begin{eqnarray}}
\def\eea{\end{eqnarray}}
\def\nn{\nonumber}

\begin{document}

\markboth{J. Peralta-Ramos and Gasta\~ao Krein}
{$\eta(T)/s(T)$ and collective flow in chiral hydrodynamics}

%
\catchline{}{}{}{}{}
%

\title{$\eta(T)/s (T)$ AND COLLECTIVE FLOW IN CHIRAL HYDRODYNAMICS}

\author{J. PERALTA-RAMOS}

\address{Instituto de F\'{\i}sica Te\'orica, Universidade Estadual Paulista\\
Rua Dr. Bento Teobaldo Ferraz, 271 - Bloco II, 01140-070 S\~ao Paulo, SP - Brazil\\
Present address: Departamento de F{\i}sica, Facultad de Ciencias Exactas y Naturales, Universidad de Buenos Aires,  
Ciudad Universitaria,  Buenos Aires 1428, Argentina.\\
jperalta@df.uba.ar}

\author{GAST\~AO KREIN}

\address{Instituto de F\'{\i}sica Te\'orica, Universidade Estadual Paulista\\
Rua Dr. Bento Teobaldo Ferraz, 271 - Bloco II, 01140-070 S\~ao Paulo, SP - Brazil \\
gkrein@ift.unesp.br
}

\maketitle

\begin{history}
\received{Day Month Year}
\revised{Day Month Year}
\end{history}

\begin{abstract}
Using relativistic conformal hydrodynamics coupled to the linear $\sigma$ model 
we study the evolution of matter created in heavy--ion collisions. We focus the
study on the influence of the dynamics of the chiral fields on the charged-hadron 
elliptic flow $v_2$ for a temperature--independent as well as for a temperature--dependent  
$\eta/s$ that is calculated from kinetic theory. We find that $v_2$ is not very sensitive 
to the coupling of chiral fields to the hydrodynamic evolution, but the temperature 
dependence of $\eta/s$ plays a much bigger role on this observable.

\end{abstract}

\ccode{PACS numbers: 25.75.-q, 24.10.Nz, 24.85.+p}

\section{Introduction}	

In any attempt of describing experimental results of ultrarelativistic heavy 
ion collisions at the Relativistic Heavy Ion Collider (RHIC) and the Large Hadron 
Collider (LHC) via a viscous fluid dynamic model, crucial physics input reflecting 
the properties of the flowing nuclear matter must be supplemented, like the equation 
of state (EOS) and transport coefficients such as the shear ($\eta$) and bulk 
($\zeta$) viscosities \cite{REVS}. 

The EOS of nuclear matter created at RHIC and LHC can be obtained from Lattice 
QCD simulations and is fairly known in the low chemical potential regime of heavy 
ion collisions. In contrast, the temperature dependence of the transport coefficients 
describing the flow of this matter is not precisely known. Although it is expected that 
for the Quark--Gluon Plasma (QGP) $\eta$ and $\zeta$ will depend strongly on
temperature, usually in hydrodynamic simulations temperature--independent values 
for these coefficients are assumed throughout the entire evolution. The impact of the 
temperature dependence of transport coefficients on momentum anisotropies as 
obtained from fluid dynamic models has been investigated only 
recently\cite{Nagle,shen,niemi}.

The comparison of hydrodynamic models to data would allow, in principle, to pinpoint 
the location of the QCD phase transition or of a crossover from hadronic matter to the 
QGP. In this context, in addition to the hydrodynamic degrees of freedom related to 
energy-momentum conservation, degrees of freedom associated with order parameters of 
broken continuous symmetries must be considered as well since they are all coupled to 
each other (see e.g. Refs.\cite{paech,krein}). 

We study, in the context of second--order dissipative relativistic fluid 
dynamics -- see Refs.\cite{REVS,PRC,luzum,hyd1,hyd2} for details, the influence 
of the long--wavelength modes of chiral fields on the expansion of the fireball 
created in Au$+$Au collisions at $\sqrt{s_{NN}} = 200$~GeV. In particular, we aim at 
quantifying the effect of the coupled evolution of chiral degrees of freedom on the 
flow asymmetry characterized by $v_2$, when shear finite viscosity is taken into 
account within a simple microscopic model for the chiral condensate. 

\section{Conformal fluid dynamics coupled to the linear $\sigma$ model}

In this section we describe the chiral--hydrodynamic model used to compute $v_2$. The quark 
fluid is described by second--order conformal fluid dynamics\cite{hyd1,hyd2}, while the 
chiral dynamics is obtained from the linear $\sigma$ model (LSM). Further details on the 
model and its numerical implementation can be found in Ref.\cite{krein}.

The classical equations of motion of the LSM are given by 
\begin{equation}
  D_\mu D^\mu \phi_a + \frac{\delta U}{\delta \phi^a} = -g\rho_a 
\label{eomchiral}
\end{equation}
where $\phi=(\sigma,\vec{\pi})$ and 
\begin{equation}
  \rho_a = g \phi_a d_q \int \frac{d^3 p}{(2\pi)^3} \frac{1}{\sqrt{p^2 + g^2 \sum_a \phi_a^2 }} f_0(p^\mu,x^\mu) 
\label{rhosrhops}
\end{equation}
$f_0(p^\mu,x^\mu)$ is the Fermi--Dirac function and $U$ is the potential exhibiting chiral symmetry breaking. The quarks act as a thermal bath, leading to an effective potential 
\begin{equation}\nonumber
 V_e (\phi_a,T) = U(\phi_a) - d_q T \int \frac{d^3p}{(2\pi)^3}\ln [1+e^{-E/T}]
\end{equation}
where $d_q = 24$ and $E=\sqrt{p^2+m_q^2}$ with $m_q^2 = g^2\sum_a \phi_a^2$. 

The classical equations of motion of the chiral fields are solved selfconsistently together with the conservation equations for the quark fluid  
\begin{equation}
\begin{split}
D\epsilon &= -(\epsilon+p)\nabla_\mu u^\mu + \Pi^{\mu\nu}\sigma_{\mu\nu} + g (\rho_s D \sigma + \vec{\rho}_{ps}\cdot D\vec{\pi}) \\
(\epsilon+p)Du^i &= \frac{1}{3}(g^{ij}\partial_j \epsilon - u^i u^\alpha \partial_\alpha \epsilon) - \Delta^i_\alpha D_\beta \Pi^{\alpha\beta} + g (\rho_s \nabla^i \sigma + \vec{\rho}_{ps}\cdot \nabla^i\vec{\pi}) ~. 
\end{split}
\label{conseq}
\end{equation}
where $(\rho_s,\vec{\rho}_{ps})=\rho_a$, $D=u_\mu D^\mu$ is the comoving time derivative,   
$\nabla_\mu = \Delta_{\mu\alpha}D^\alpha$ is the spatial gradient, and $\sigma_{\mu\nu}$ is shear tensor. 
The evolution equation for $\Pi^{\mu\nu}$ contains terms up to second order in velocity gradients and reads
\begin{eqnarray}
\partial_\tau \Pi^{i\alpha} &=& -\frac{4}{3u^\tau}\Pi^{i\alpha}\nabla_\mu u^\mu - \frac{1}{\tau_\pi u^\tau}\Pi^{i\alpha} + \frac{\eta}{\tau_\pi u^\tau} \sigma^{i\alpha} -\\ \nonumber
& -& \frac{\lambda_1}{2\tau_\pi \eta^2 u^\tau}\Pi^{<i}_\mu \Pi^{\alpha> \mu} - \frac{u^i\Pi^\alpha_\mu + u^\alpha \Pi^i_\mu}{u^\tau}Du^\mu  -\frac{u^j}{u^\tau}\partial_j \Pi^{i\alpha} 
\end{eqnarray}
where $\eta$ is the  shear viscosity and $(\tau_\pi,\lambda_1)$ are second--order transport coefficients. 
From Eqs. \ref{eomchiral,conseq} it is seen that the evolution of the chiral fields affects the evolution of the 
quark fluid through the sources in the energy--momentum conservation equations; 
in turn, the quark fluid affects the evolution of the chiral fields through the 
densities ${\rho}_s$ and $\vec\rho_{ps}$, that depend on the local values of the fluid dynamic variables $T$ and $u^\mu$. 

In order to compute the temperature dependence of the shear viscosity $\eta(T)$ in the LSM, we adapt the method described in Ref.\cite{sas} and employ the linearized Boltzmann equation in the relaxation time approximation (see Ref.\cite{krein}). 
The shear viscosity is given by
\begin{equation}
\eta = \frac{4\tau}{5T}\int  \frac{d^3p}{(2\pi)^3} \frac{p^4}{E^2}f_0 (1-f_0)
\end{equation}
where $\tau=\tau(T)$ is the collision time calculated from the averaged cross sections $\bar{\sigma}$ 
for quark--quark and quark--antiquark scattering processes including $1/N_c$ next to leading order corrections as
\bea
\tau^{-1} &=& 6 f_0 \Bigl( \bar{\sigma}_{uu\rightarrow uu} + \bar{\sigma}_{ud\rightarrow ud}
+ \bar{\sigma}_{u\bar{u}\rightarrow u\bar{u}} + \bar{\sigma}_{u\bar{u}\rightarrow d\bar{d}}
\nn\\
&& + \, \bar{\sigma}_{u\bar{d}\rightarrow u\bar{d}}\Bigr) ~~.
\eea
We refer the reader to Ref.\cite{sas} for details on the
calculation of the $\bar{\sigma}'s$ 

\begin{figure}[pb]
\centerline{\psfig{file=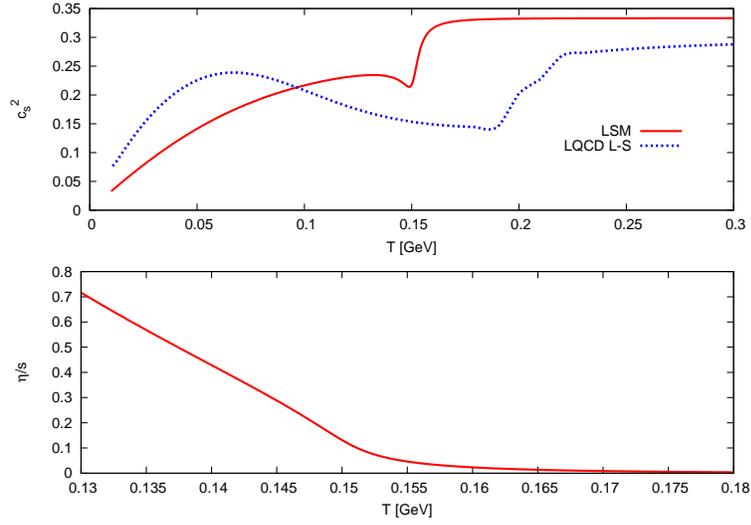,width=10cm}}
\vspace*{8pt}
\caption{Speed of sound of the LSM and that of the model EOS of Laine and Schr\"{o}der$^{12}$ connecting a high-order weak-coupling perturbative QCD calculation to a hadron resonance gas through a crossover ({\it upper panel}); and $\eta/s(T)$ in the LSM with $g=3.2$ ({\it lower panel}). }
\label{cs}
\end{figure}

\begin{figure}[pb]
\centerline{\psfig{file=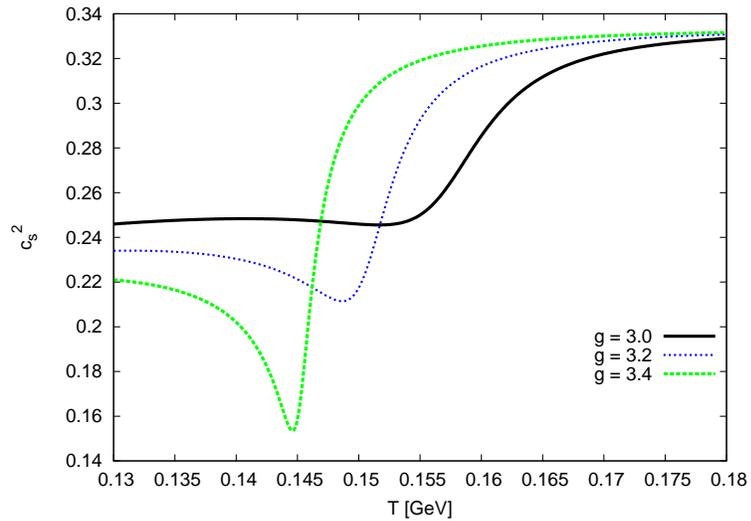,width=10cm}}
\vspace*{8pt}
\caption{Square of the sound speed $c_s^2$ of the LSM as a function of 
temperature, for three values of the chiral coupling: $g=3,3.2,3.4$.}
\label{cg}
\end{figure}

The speed of sound $c_s^2$ and $\eta/s$ of the LSM with $g=3.2$ is shown in Fig. \ref{cs}, together with the $c_s^2$ corresponding to Lattice QCD\cite{laine} (shown for comparison). The value of $g=3.2$ is chosen because it leads to a smooth crossover with a drop in $c_s^2$ near the critical temperature\footnote{Strictly speaking, there is no $T_c$ since we are dealing with a crossover and not with a phase transition. However, we will still refer to a $T_c$ as an approximate critical temperature around which $c_s^2$ varies rapidly} $T_c$ which is comparable to the one obtained from Lattice QCD simulations. For larger (smaller) values of $g$, the drop in $c_s^2$ near $T_c$ is significantly larger (smaller), as shown in Fig.~\ref{cg}. It is seen that decreasing the value of $g$ leads to a softening of the crossover and an increase in $T_c$, while increasing it leads to a decrease in $T_c$ and a steeper variation of $c_s^2$ with temperature. We note that a first order phase transition is obtained if $g\simeq 3.8$, but we will not consider this case since Lattice QCD simulations\cite{laine} indicate the QGP--hadron transition is a smooth crossover (see Fig.~\ref{cs}).

For the second--order transport coefficients we take\cite{krein}: $\tau_\pi = 2(2-\ln 2)\frac{\eta}{sT}$ and $\lambda_1=\frac{\eta}{2\pi T}$ corresponding to the $\mathcal{N}=4$ Super Yang--Mills theory. We use a 13 fm $\times$ 13 fm transverse plane, set the impact parameter to  $b=7$ fm and the initialization time to $\tau_0 = 1$ fm/c. As initial conditions we use $u^x=u^y=0$ and $\Pi^{xx}=\Pi^{xy}=\Pi^{yy}=0$. These are typical values for the parameters used as input in hydrodynamic simulations of heavy ion collisions at RHIC, and were used in Refs. \cite{PRC,krein}.
The initial energy density profile is obtained from Glauber's model, with a temperature $T_i=330$ MeV at the center of the fireball. As a reasonable ansatz we take $\vec{\pi}(\tau_0,\vec{r})=0$ and $\sigma(\tau_0,\vec{r}) = f_\pi [1-e^{-(r/r_0)^2}]$ with $r_0 = 9$ fm as initial values for the chiral fields. We use the isothermal Cooper-Frye freeze--out prescription with a freeze-out temperature $T_F = 130$ MeV.

\section{Charged--hadron elliptic flow}

We will now discuss the results obtained for the charged--hadron elliptic flow $v_2$ computed 
from the chiral--hydrodynamic model described above. 

\begin{figure}[pb]
\centerline{\psfig{file=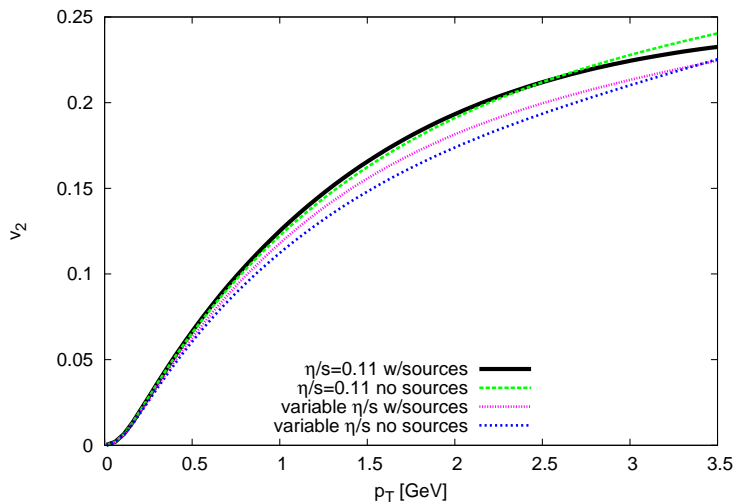,width=10cm}}
\vspace*{8pt}
\caption{Elliptic flow of charged--hadrons calculated taking or not taking into 
account the source terms in the hydrodynamic equations, for either a 
temperature--dependent or a temperature--independent $\eta/s$ (taken as the 
averaged value throughout the evolution of the fireball). The 
value of the coupling constant is $g=3.2$.}
\label{v2}
\end{figure}

Fig. \ref{v2} shows $v_2$ calculated taking or not taking into 
account the source terms in the hydrodynamic equations, for either the temperature--dependent computed from kinetic theory or a temperature--independent $\eta/s$ which is chosen to be the averaged value throughout the evolution of the fireball. 
The results shown were obtained with a value $g=3.2$ in the LSM, which corresponds to a smooth crossover between hadronic matter and the QGP as suggested by Lattice QCD calculations 
(see e.g. Ref.\cite{laine}).

\begin{figure}[pb]
\centerline{\psfig{file=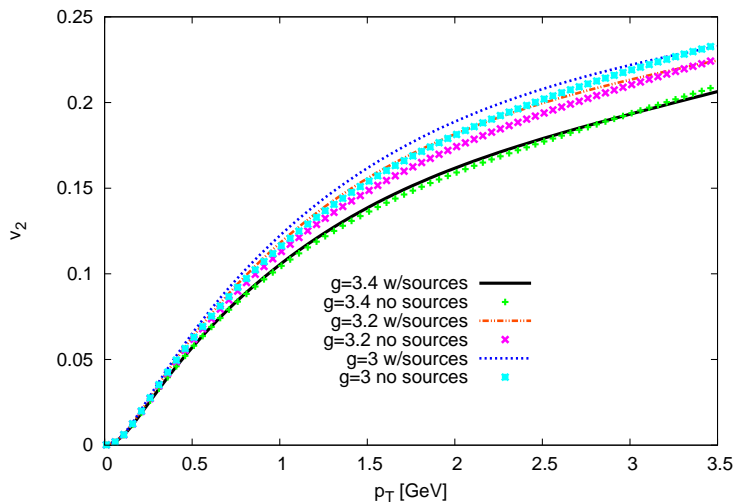,width=10cm}}
\vspace*{8pt}
\caption{Elliptic flow of charged--hadrons calculated taking or not 
taking into account the source terms in the hydrodynamic equations, for $g=3,3.2,3.4$ 
and a temperature--dependent $\eta/s$.}
\label{v2g}
\end{figure}

It is seen from Fig. \ref{v2} that both for a temperature--dependent and a temperature--independent $\eta/s$, 
the elliptic flow does not depend strongly on the chiral sources. More precisely, we see that for a temperature--independent 
$\eta/s$, $v_2$ turns out to be practically independent of the dynamics of the chiral sources for 
$p_T~<~3$~GeV.
In contrast, 
for the temperature--dependent $\eta/s$ there are small but visible differences in the elliptic flow 
calculated with or without the chiral sources. Specifically, $v_2$ is slightly larger at low $p_T$ and saturates at a smaller value of 
$p_T$ when the chiral fields are taken into account as sources for the hydrodynamic
variables.

In order to determine the impact of the value of $g$ on the behavior of $v_2$ with $p_T$, we also calculated $v_2$ for a temperature--dependent $\eta/s$ 
and different values of $g$ close to $g=3.2$. The results are shown in Fig.~\ref{v2g}. It is seen that when the value of $g$ increases, $v_2$ 
decreases (this is due to the fact that for larger values of $g$, we find that the average value of $\eta/s$ 
throughout the hydrodynamic evolution becomes smaller). Comparing the values of $v_2$ calculated taking or not taking into account the chiral fields as sources for different values of $g$, it is seen that the influence of chiral dynamics on $v_2$ becomes smaller as $g$ increases. This supports our earlier conclusion that the evolution of the chiral fields have only a small impact on $v_2$.

It seems clear that there are difficulties in extracting (an average 
value of) $\eta/s$ from data on $v_2$. Our results show that uncertainties associated 
with the dependence of $\eta/s$ on temperature lead to appreciable changes in the curve 
of $v_2$ versus $p_T$, much larger than the changes stemming from the coupling of fluid 
dynamic variables to evolving classical chiral fields. The uncertainty associated with the
temperature--dependence of $\eta/s$ should be added to the theoretical uncertainty that 
comes e.g. from the initial conditions (for example using Color Glass Condensate or 
Glauber initial conditions) and the freeze--out process, that according to recent studies
add up to an overall uncertainty which can be roughly 
estimated in $0.1$~(see e.g. Refs.\cite{REVS,PRC,luzum}).

\section{Conclusions}

We have found that the values of $v_2$ do not depend strongly on the evolution 
of the chiral fields. Specifically, for a temperature--independent $\eta/s$ this 
dependence is negligible for $p_T < 3$ GeV, while for a temperature--dependent $\eta/s$ 
it is appreciable but still small even at small $p_T$.  

In line with the results of Refs.\cite{Nagle,niemi,shen}, our results show that not 
knowing precisely the temperature--dependence of $\eta/s$ leads to significant uncertainties 
in attempts of extracting this ratio from data on $v_2$, in addition to the uncertainties 
that stem from the initial conditions and the freeze--out process, among others sources.  

It is worth noting that despite the coupling of chiral sources to the hydrodynamic 
evolution would add further uncertainties to the values of $\eta/s$ extracted from data, 
they are not very big and do not affect in a significant way the possible extraction of 
$\eta/s$ values from charged--hadron elliptic flow data. 

Future work includes improvements such as the consideration of bulk 
viscosity in the hydrodynamic equations and fluctuations of the chiral 
fields. This latter effect would act as noise sources in the classical 
equations of motion\cite{Farias:2007xc,CassolSeewald:2007ru,Nahrgang:2011mg}.

\section*{Acknowledgments}

We thank Esteban Calzetta for useful discussions. 
This work was partially funded by CNPq and FAPESP (Brazilian agencies), and CONICET (Argentina).

\end{document}